\documentclass[12pt,preprint]{aastex}

\newcommand{\rxj}{RX~J$1856.5-$3754}
\newcommand{\pks}{PKS~$2155 -$304}

\slugcomment{To appear in the Astrophysical Journal}

\shorttitle{Is RX J185635-375 a Quark Star?}
\shortauthors{Authors et al}

\begin{document}

\title{Is RXJ1856.5-3754 a Quark Star?}
\author{Jeremy~J.~Drake\altaffilmark{1}, 
Herman~L.~Marshall\altaffilmark{2}, 
Stefan~Dreizler\altaffilmark{3}, 
Peter~E.~Freeman\altaffilmark{1}, 
Antonella~Fruscione\altaffilmark{1}, 
Michael~Juda\altaffilmark{1}, 
Vinay~Kashyap\altaffilmark{1}, 
Fabrizio~Nicastro\altaffilmark{1}, 
Deron O.~Pease\altaffilmark{1},
Bradford~J.~Wargelin\altaffilmark{1}, 
Klaus~Werner\altaffilmark{3}
}
\affil{$^1$Smithsonian Astrophysical Observatory,
MS-3, 60 Garden Street, Cambridge, MA 02138}
\affil{$^2$MIT Center for Space Research, Cambridge, MA 02139}
\affil{$^3$Institut f\"ur Astronomie und Astrophysik, Universit\"at
T\"ubingen, Sand 1, 72076 T\"ubingen, Germany}

\begin{abstract}
Deep Chandra LETG+HRC-S observations of the isolated neutron star
candidate \rxj\ have been analysed to search for metallic and
resonance cyclotron spectral features and for pulsation behaviour.  As
found from earlier observations, the X-ray spectrum is well-represented by a
$\sim 60$~eV ($7\times 10^5$~K) blackbody.
No unequivocal evidence of 
spectral line or edge features has been found, arguing against
metal-dominated models.
The data contain no evidence for pulsation and we place a 99\%\
confidence upper limit of 2.7\%\ on the unaccelerated pulse fraction 
over a wide 
frequency range from $10^{-4}$ to 100 Hz.  We argue that the derived
interstellar medium neutral hydrogen column density of $8\times
10^{19}\leq N_H \leq 1.1\times 10^{20}$ cm$^{-2}$ favours the larger
distance from two recent HST parallax analyses, placing \rxj\ at $\sim
140$~pc instead of $\sim 60$~pc, and in the outskirts of the R~CrA
dark molecular cloud.  That such a comparatively rare region of high
ISM density is precisely where an isolated neutron star re-heated by
accretion of interstellar matter would be expected is either entirely
coincidental, or current theoretical arguments excluding this scenario
for \rxj\ are premature.  Taken at face value, the combined
observational evidence---a lack 
of spectral and temporal features, and an implied radius
$R_\infty=3.8$-8.2~km that is too
small for current neutron star models---points to a more
compact object, such as allowed for quark matter equations of state.

\end{abstract}


\section{Introduction}
\label{s:intro}

The structure and evolution of neutron stars depends on the properties
of matter at nuclear and supranuclear densities.  Such conditions are
not achievable in terrestrial laboratories and the theoretical
description of such superdense matter remains uncertain.  It has
therefore been hoped that neutron stars, and in particular those that
are isolated and not complicated by strong accretion or magnetospheric
signatures, might provide some empirical insights: observations of
their masses, radii and cooling characteristics could, in principle,
provide useful constraints for the equation of state (EOS) of dense
matter (e.g.\ Lattimer \& Prakash 2001 and references therein).

In a relatively brief period of $10^6$-$10^7$~yr, a hot, isolated
neutron star (INS) born in a supernova explosion can cool, cease
pulsar activity and become essentially inactive (see, e.g.,
the review of Treves et al.\ 2000).  Of the estimated $10^8$-$10^9$
isolated neutron stars thought to inhabit the Galaxy, only a tiny
fraction are therefore expected to be sufficiently young to remain hot
and visible in X-rays.  One possible mechanism capable of sustaining thermal
X-ray emission in an older INS is accretion of material from the ISM.
To date, only a handful of these older INS
candidates have been found.

The soft X-ray source \rxj\ discovered by Walter, Wolk \& Neuh\"auser
(1996) is the brightest, and probably the closest (Kaplan, van
Kerkwijk \& Anderson 2002) of the INS candidates.  It was identified
with a very faint 
($V\simeq25.6$) optical counterpart by Walter \& Matthews (1997), was
found to have an optical flux about a factor of 2-3 higher than that
predicted by the Rayleigh-Jeans tail of the $\sim 55$~eV blackbody
spectrum that represents the low resolution ROSAT PSPC spectrum (e.g.\
Walter et al.\ 1996; Campana et al.\ 1997, Pons et al.\ 2001), and
lies in the line-of-sight toward the dark molecular cloud R CrA.
However, the exact nature of \rxj\ remains unknown---whether it is a
fairly young, cooling INS, perhaps undetected as a pulsar because of
unfortunate beam alignment, or an older object reheated by ISM
accretion.

Very Large Telescope (VLT) observations have recently revealed an 
H$\alpha$ nebula round \rxj\ and a blackbody spectrum through the
UV-optical range 
(van Kerkwijk \& Kulkarni 2001a,b).  
Hubble Space Telescope (HST) astrometry was used by Walter (2001) and
Kaplan et al.\ (2002) to estimate a parallax and proper motion, but
with conflicting results ($16.5\pm 2.3$~mas vs.\ $7\pm 2$~mas).  Walter
(2001) argued that the proper motion points to the Sco-Cen OB
association and an age of $\sim 10^6$~yr.  However, Pons et al.\
(2001) failed to detect the expected pulsation signature 
in ROSAT and ASCA data.  Modelling X-ray, EUV, UV and optical
spectra using the Walter (2001) parallax and assuming a metal-dominated
atmosphere resulted in stellar radii too small for current EOS and
smaller than the Schwarzschild radius for a canonical $1.4 M_\odot$
star, leading them to conclude that the surface temperature
distribution could be inhomogeneous.  In contrast, the same 
atmospheric analysis using the larger distance of
Kaplan et al.\ (2002) would yield a radius consistent with current theory.
If the atmosphere is indeed metal-dominated, then line and edge
features should be visible in high resolution X-ray spectra.

\rxj\ was observed in 2000 March for 55ks by the Chandra X-ray
Observatory Low Energy Transmission Grating (LETG) and the High
Resolution Camera spectroscopic microchannel plate detector array
(HRC-S) to look for spectral features and pulsar activity.  Within
relatively large statistical uncertainties, Burwitz et al.\ (2001)
failed to detect either significant departures from a blackbody
spectrum or pulsations in these data.  The prospect of important
scientific gains from a longer observation prompted, at the request of
different researchers, the undertaking of a very recent set of Chandra
observations of \rxj\ under director's discretionary time using
LETG+HRC-S.  A period search using these data by Ransom, Gaensler \&
Slane (2002) has already placed an upper limit on a pulse fraction of
4.5\%.  This {\em Letter} presents spectroscopic and independent
timing analyses of these data.

\section{Observations and Data Reduction}
\label{s:obs}

The observations analysed in this paper 
Pipeline-processed (CXC software version 6.3.1) photon event lists
were reduced and analysed using the CIAO software package version 2.2,
and independently using custom IDL\footnote{Interactive Data Language,
Research Systems Inc.}  software.  Processing included extra
pulse-height filtering to reduce background (Wargelin et al., in
preparation) and barycentric correction of event times.  Dispersed
photon events were extracted using the now standard ``bow tie''
window, and a small circular region (31-pixel radius\footnote{Our 31
pixel radius extraction region for the 0th order source corresponds to
an encircled energy fraction of about $(92\pm 3)$\%, which combines
with an average deadtime of 0.59\%\ to yield a correction factor for a
$2\pi$ steradian aperture of 1.09.}) was used to extract the 0th order
events.  We note that an extraneous bright feature (likely a ghost
image of a bright off-axis source) appeared on the +1 order outer
HRC-S plate which introduces spurious features into the extracted
spectrum and background near 110 \AA.  We applied corrections to
extracted event times to ameliorate an HRC electronics problem that
assigns time tags for each event to the event that immediately
follows.  If every event were telemetered to the ground, correct times
could be easily reassigned, but because of the high HRC background
rate and a telemetry limit of 184 events/sec, only {\it valid} events,
which make up approximately 45\% of the {\it total} events in these
observations, are recorded in data received on the ground.  We have
reassigned the time of every valid event to the preceding valid event
during ground processing; the event times will then be correct about
$N_{valid}/N_{total}$ of the time.  One can place an upper limit of
$\delta t$ on the time errors associated with this process, however,
simply by excluding events with a time-shift of more than $\delta t$;
the average timing error of such events will, of course, be less than
$\delta t$.  However, if $\delta t$ is too small, too many events get
excluded to perform sensitive timing studies.  We have adopted a
$\delta t$ of 2~ms, retaining 20\% of the
original counts in the corrected data set.

\section{Count Rates and Timing Analysis}
\label{s:timing}

Count rates were derived for each observation segment from event lists
filtered to exclude times of high background and
telemetry saturation, when the telemetered valid event rate exceeded
184 count~s$^{-1}$, and to exclude high pulse-height events that arise
entirely from background.  These rates corresponding to 0th order only
are listed in Table~\ref{t:obs}.

The count rates for the different observation segments are
statistically consistent, though the 2000 March observation (113) rate
of $0.2195\pm 0.0020$~count~s$^{-1}$ lies 1.9\%\ and $1.9\sigma$ above
the rate for the combined 2001 October series rate of $0.2155\pm
0.0008$.  Such deviations will be obtained by chance about 3\%\ of the
time when the count rates do not vary; it is thus 
more likely attributable to quantum efficiency (QE) variations in the
detector on small scales.  Indeed, Pons et al.\ (2001) note two ROSAT
HRI observations obtained 3~yr apart that are consistent to 1\%.  The
fluxes corresponding to HRI and PSPC count rates are higher than that
obtained from our Chandra data by 20\%\ and 30\%, respectively.  As
remarked by Burwitz et al.\ (2001), these differences are likely
attributable to absolute calibration uncertainties.  We have also
examined the ASCA SIS observation described Pons et al.\ (2001) and
find fluxes 30-40~\%\ lower than obtained by Chandra for the
20-30~\AA\ range, but in agreement with Chandra shortward of 20~\AA .
The EUVE DS count rate of Pons et al.\ (2001) is also consistent with
the Chandra observation within allowed uncertainties.

Our search for pulsations used three different techniques, none of
which found any evidence for significant variability.  In contrast to
Ransom et al.\ (2002), we did not include a deceleration term; this
will be discussed below. 

We applied the Bayesian
method of Gregory \& Loredo (1992) to both the time-corrected and
uncorrected 0th order event lists.  The method tests for variability by
comparing the fits of periodic stepwise models to the data with the
fit of a constant model.  The odds favoring variability (based on
eqn.\ 5.28 of Gregory \& Loredo 1992, with a maximum number of steps
$m_{\rm max}$ = 12 and limiting angular frequencies $\omega_{\rm lo}$
and $\omega_{\rm hi}$ equal to 10$^{-4}$ and 10$^3$, respectively)
were found to be $1.45\times 10^{-4}$ for the whole data set and
$3.75\times 10^{-3}$ for the data filtered on $\delta t=2$~ms, both to
be compared with an odds value of $10^2$ needed for a confident
pulsation or variability detection.

Our second method employed an FFT analysis applied to the combined 0th
and 1st order events, followed by a likelihood
ratio test (LRT) to determine limits on the pulse fraction.  The FFT
power distribution was consistent with shot noise.  For the LRT of
a given period, $P$, the data were binned into $N$ phase bins, giving
$n_i$ counts in each.  The source model was $y = A + f
\cos(\phi_i+\phi_0)$ where $\phi_i$ is the phase of bin $i$ and
$\phi_0$ is the phase of the pulse and
$f/A$ is the pulse fraction.  The likelihood equations were then
solved for $A$ and the process applied to $> 500$ frequencies where
the FFT power exceeded a critical level in the frequency range
0.001-50~Hz.  By including also the dispersed events we
improved the signal-to-noise of the result by a factor of 1.47 compared
to using 0th order alone and could obtain a pulse fraction limit
lower than the value of 4.5\%\ obtained by Ransom et al.\
(2002) in their unaccelerated search.  A pulse fraction upper limit
(99\%\ confidence) of 
2.7\%\ was derived applying our likelihood ratio method using all
data, including the dispersed events with $1 < \lambda < 70$ \AA .
Taking only the events limited by $\delta t < 2$ ms, the pulse
fraction limit is 10\%.

Thirdly, we computed the Lomb-Scargle periodogram for both the
time-corrected and uncorrected photon arrival time differences in the
frequency range 0.01-1~Hz for the events in ObsID's 3380 3381, 3382
and 3399.  Again, no significant peaks were present.

The assumption of a negligible deceleration term in our period search
restricts the range of periods and dipole magnetic field strengths for
which our search is valid.  The coherence limit for phase slippage by
10~\%\ over the duration of the 2001 October observations implies that
our result is valid for a magnetic field upper limit $B < 2.3\times 10^{13}
P^{3/2}$~G for period $P$~s (e.g.\ Shapiro \& Teukolsky 1983).  As
noted by Ransom et al.\ (2002), this 
range would exclude very young and energetic neutron stars, such
as the Crab and Vela pulsars, though most of these younger objects
are also conspicuously strong radio pulsars.  All anomalous X-ray pulsars 
would lie within our sensitivity limit range.  \rxj\ is also most
unlikely to be an extremely young object based on its modest 
temperature and luminosity, which are consistent with an object of age
$\sim 10^5$~yr on canonical cooling curves (e.g.\ Tsuruta 1997).

\section{Spectral Analysis and Model Parameter Estimation}
\label{s:spectral}

\begin{figure}
\plotone{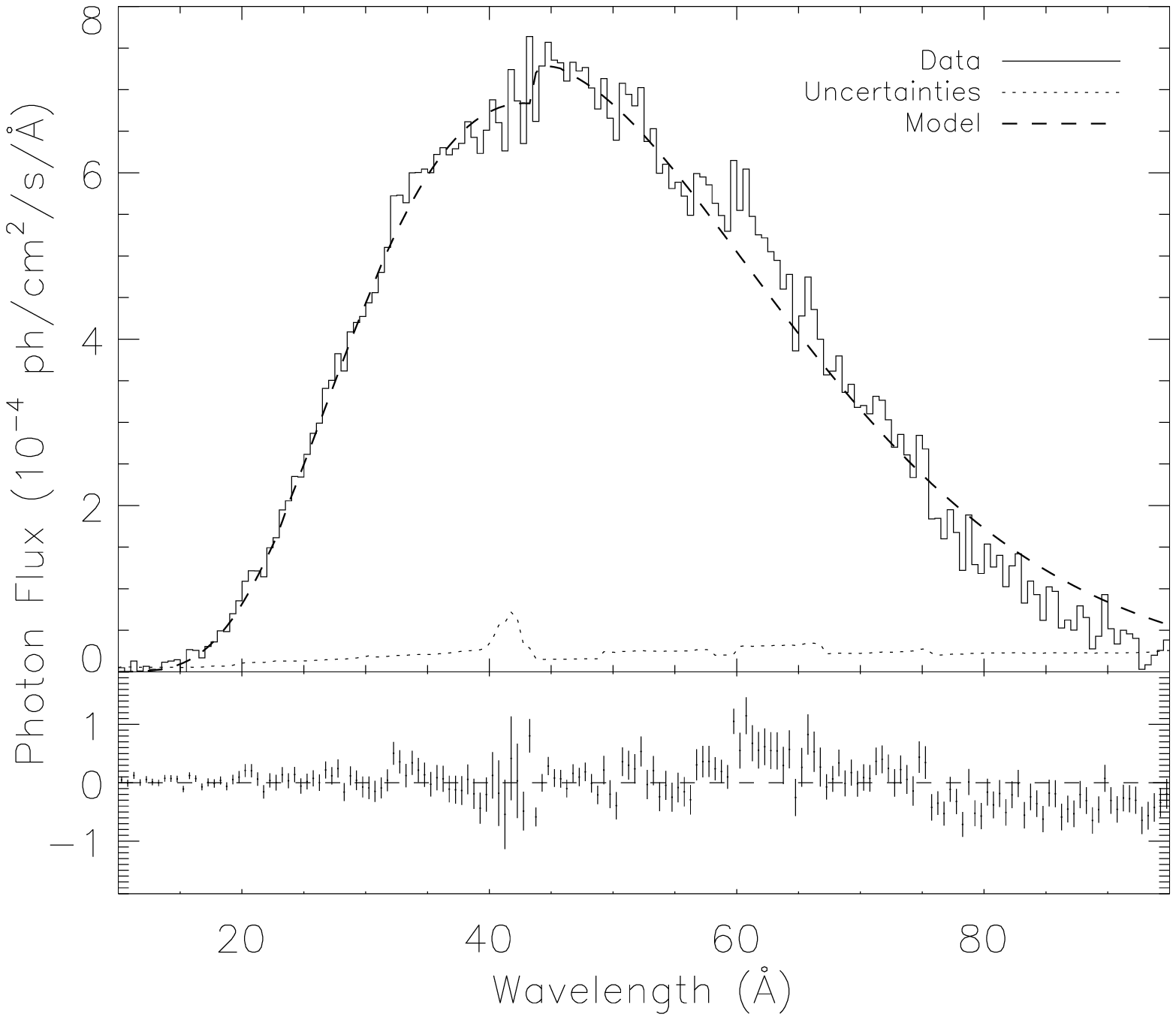}
\caption{
The combined positive and negative order spectra of \rxj\
binned at 0.5~\AA\ intervals shown with the best fit blackbody model with
parameters corresponding to method (2) in \S\ref{s:spectral} and
residuals (observations$-$model).  The deviations from this model are
consistent with Poisson statistics after allowing for 
calibration uncertainties at the C K-edge and over broader wavelength
intervals.  The apparent edge at 60 \AA\ results primarily from one 
of the HRC-S plate gap boundaries and small residual QE differences
between positive and relative negative order outer plates.
\label{f:spectra}}
\end{figure}

Spectral analysis in the form of model parameter estimation was
undertaken using the CIAO/Sherpa fitting engine and independently
using specially-written IDL software.  Cursory inspection of the
spectrum leads immediately to the conclusion that there are no obvious
features indicative of absorption lines or edges.  We found that blackbody
models represent the high resolution spectra well, in agreement with
earlier studies.  We modelled +1 and $-1$ orders both separately and
simultaneously, and added together; results from these different
approaches were statistically indistinguishable.  Representative
results of model fits and residuals are illustrated in
Figure~\ref{f:spectra}.  Two sets of best-fit parameters were obtained
from independent analyses that invoked (1) the existing first and higher order
CXC calibration\footnote{Version dated 2000 October 31;
http://asc.harvard.edu/cal/Links/Letg/User} 
and (2) the same first order effective area with higher orders 
modified slightly to improve 
model fits to sources with power law spectra (Marshall et al., in
preparation).  Both sets are consistent with the
results of Burwitz et al.\ (2001) based on the 2000 March observation alone. 
Parameters and $1 \sigma$ statistical uncertainties for best-fit
models were: (1) $T=61.2\pm 0.3$~eV; $N_H=(1.10\pm 0.02)\times
10^{20}$~cm$^{-2}$; 
X-ray luminosity $(2.96\pm0.03)\times 10^{31}D_{100}^2$~erg~s$^{-1}$, where
$D_{100}$ is the distance in units of 100~pc; (2) $T=61.1\pm 0.3$~eV;
$N_H=(0.81\pm 0.02)\times 10^{20}$~cm$^{-2}$; X-ray luminosity
$(3.16\pm 0.03)\times 
10^{31}D_{100}^2$~erg~s$^{-1}$.
Parameters producing minima in the $\chi^2$ test statistic
were not sensitive to the exact binning adopted, 
though of course the reduced $\chi^2$
values were: values ranged from 0.94 for data binned to a
signal-to-noise ratio of $S/N=10$, to 1.7 for $S/N=30$.  The latter
value is dominated by residual effective area calibration uncertainties.
To investigate the effects of these 
uncertainties, which are estimated to be about 15\%\ absolutely, 
a first order polynomial term was included in the source model 
to mimic an effective area lower by 15\%\ at 20\AA\ and higher by 15\%\
at 100~\AA , and vice-versa.  Such a slope skews the blackbody curve
and leads systematically to higher or lower temperature solutions by
about 1~eV---clearly the true temperature uncertainty is driven by this
uncertainty in the effective area.  Allowing for this uncertainty, 
we adopt a final temperature of $61.2\pm 1.0$---in agreement with, but 
much more tightly constrained than, the ROSAT temperatures derived by 
Burwitz et al.\ (2001; $63\pm 3$~eV) and Pons et al.\ (2002; $55.3\pm
5.5$).  We note that a temperature of 61~eV results in an optical flux
only 10\%\ higher than that from a 55~eV blackbody, so that the
discrepency between the observed optical flux and 
that predicted by the hot blackbody noted by Walter \& Matthews (1997)
and Pons et al.\ (2002) remains essentially unchanged.

Blackbody models were found by both Pons et al.\ (2001) and Burwitz et
al.\ (2001) to represent observed spectra better than sophisticated
model atmospheres, though a uniform temperature blackbody model was
formally excluded in X-ray-EUV-UV-optical modelling in the former
work.  However, additional cooler components were found to contribute
at most only a few percent to the observed ROSAT PSPC X-ray flux.  We
have ruled out the significant presence of additional thermal and
non-thermal emission components by trial of models combining two
blackbodies, and a blackbody component with arbitrary power laws: in
both cases the additional components were completely unconstrained and
resulted in no improvement in the goodness of fit.  The $3\sigma$
upper limit to power law flux is $5.2\times
10^{28}D_{100}^2$~erg~s$^{-1}$keV$^{-1}$ at 1~keV.  Our blackbody
models from methods (1) and (2) correspond formally to a radius over
distance ratio (angular size) of $R_\infty/D_{100}=4.12\pm
0.68$~km/100pc, where $R_\infty=R/\sqrt{1-2GM/Rc^2}$ is the
``radiation radius'' corresponding to the true radius $R$ for a star
of mass $M$, and the quoted uncertainty represents the combined
temperature determination uncertainty ($\pm 1$~eV) and the (dominant) absolute
effective area uncertainty of the LETG+HRC-S combination ($\pm 15$\% ).

In all model comparisons for binning at $S/N>10$ the residual
differences between the observed counts and the best-fit model show
systematic departures from normal statistical deviations
(Figure~\ref{f:spectra}).  Broad deviations are characterised by a
general overprediction of observed counts for $\lambda < 30$~\AA\ and
in the region 75-100~\AA\ which is dominated by higher order flux,
underprediction by an average of $\sim 10$~\%\ for the range
25-38~\AA, and deviations around the instrumental C~K edge region
40-44~\AA .  The latter results from a residual calibration error in the
HRC-S UV/Ion shield.  Other deviations could arise either as a result
of impropriety of a blackbody model for \rxj , or through calibration
errors which are currently estimated to be $\leq 15$\%\ over
broad spectral ranges and less over narrower ranges (Drake et al., in
preparation).  An apparent edge at 60~\AA\ and flux excess in the
range 60-70~\AA\ arises because of one of the 
HRC-S plate gap boundaries coincides with small residual QE differences
between positive and relative negative order outer plates.

\begin{figure}
\plotone{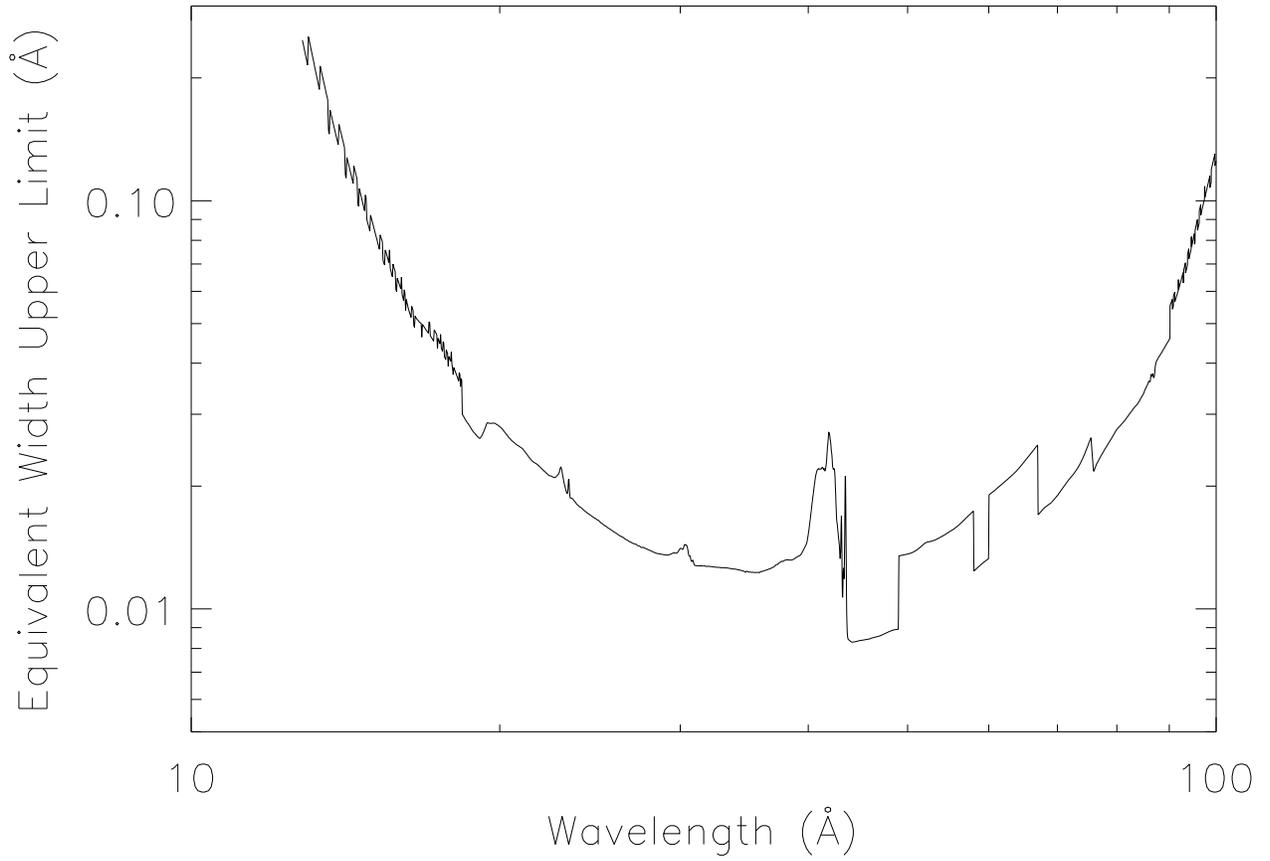}
\caption{The $3\sigma$ equivalent width upper limit to line features 
as a function of wavelength based on a convolution of the spectrum
with a triangular kernel.
\label{f:ew}}
\end{figure}

In the case of narrow line or edge features, in different combinations
of spectral order, data reduction method and binning size, we identify
possible structure in residuals at 26.5, 27.6, 34.4, 32.4, and 35.9 \AA\
(emission like features), and at 28.2, 39.1, and 86.5 \AA\ (absorption
like features), that might be tempting to attribute to the source.
However, we cannot exclude the possibility that any are chance
fluctuations at the 1\% level.  The issue is complicated by possible
calibration uncertainties on smaller scales, believed to be at a level
of about 5\%\ or less, that should be largely smoothed out by
dither.  We have also used a 60ks observation (ObsID 331) of \pks ,
currently thought to be a featureless continuum in the spectral range
we are concerned with here (Marshall et al., in preparation), as a flat
field source to aid in feature identification.  This spectrum
comprises about eight times the number of first order counts of the combined
\rxj\ data.  Moreover, we have imposed a constraint that features must
appear in both positive and negative orders in the \rxj\ spectrum.  We
examined deviations from the model on scales up to 3~\AA\ and find
only the expected normal distribution of residuals after allowing for
smooth departures resulting from calibration.  The Kolmogorov-Smirnov
test applied to deviations on different scales also revealed no
evidence for significant features these scales.  In summary, all
significant deviations in the residuals that have been found can
reasonably be explained by instrumental effects.  Equivalent width
upper limits were derived by applying counting statistics to a
convolution of the spectrum with a triangular kernel
(Figure~\ref{f:ew}).

\section{Interstellar Medium Absorption}
\label{s:ism}

The distance of 62~pc derived by Walter (2001) seems at odds with the
neutral H column density, $N_H$, of $10^{20}$~cm$^{-2}$ derived from
the Chandra spectra: measured $N_H$ values for
objects at this distance based on different techniques are
typically in the range $10^{18}$-$10^{19}$~cm$^{-2}$ (e.g.\
Fruscione et al.\  1994).  Walter (2001) indeed remarked on this, citing
reddening values $E_{B-V}$ of up to 0.1 derived by Knude \& H{\o}g
(1998) in support of the distance.  However, these reddening values show
considerable scatter at low reddening and are based only on a relatively coarse
attribution of spectral type to the stars considered.  

\begin{figure}
\plotone{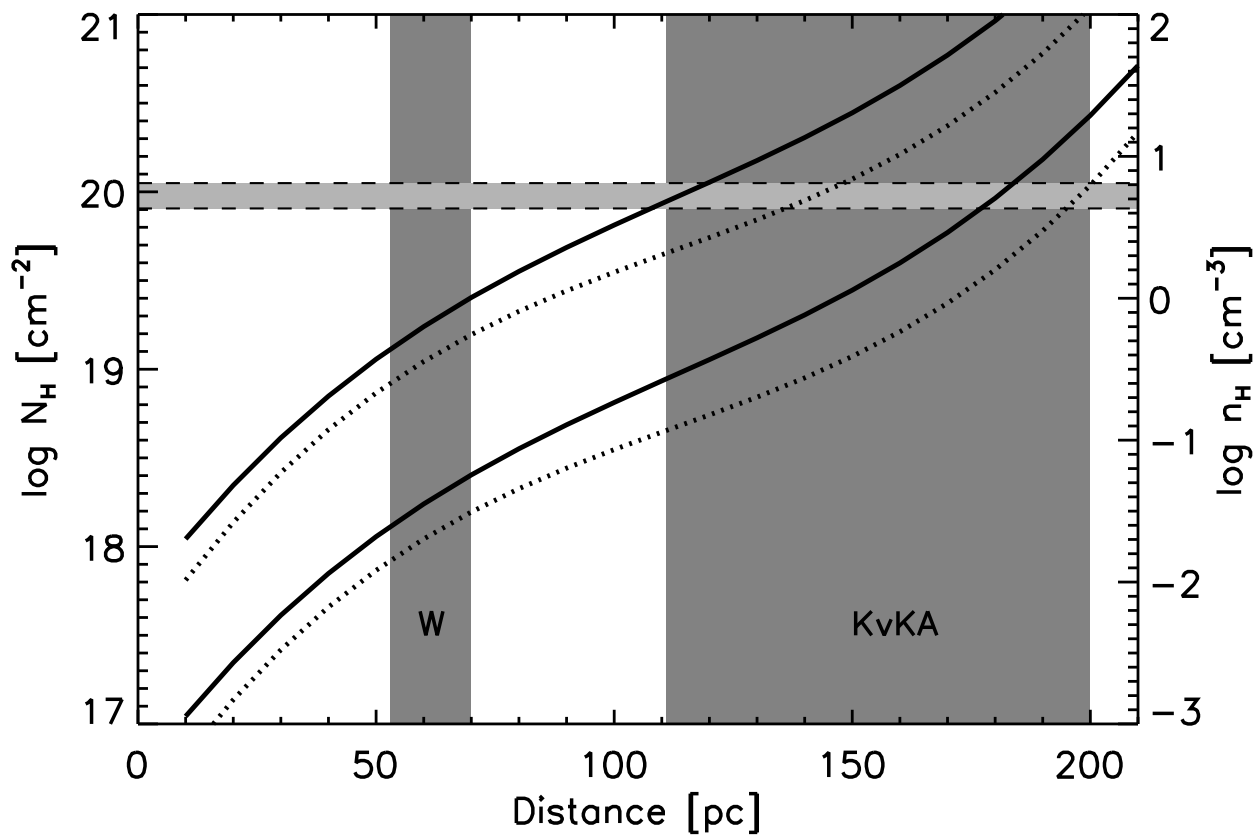}
\caption{The estimated $N_H$ (solid curves) 
and $n_H$ (dotted curves) as a function of distance in the 
line-of-sight toward \rxj .  The pairs of curves represent the likely
ranges of these quantities and correspond to deviations of factors of
3 from a smooth
locus through the results of the 3D interpolations.  Vertical
shaded regions indicate the 
distance ranges found by 
Walter (2001; W) and Kaplan et al.\ (2002; KvKA).  The horizontal stripe
illustrates the allowed range ($\pm 1\sigma$) of $N_H$
derived in this study.  
\label{f:ism}}
\end{figure}

We have estimated $N_H$ and the mean local neutral hydrogen number
density, $n_H$, in the line-of-sight toward \rxj\ as a function of
distance by spatial interpolation in the measurements compiled by
Fruscione et al.\ (1994) and Diplas \& Savage (1994) using a technique
developed by P.~Jelinsky (unpublished).  We illustrate this in
Figure~\ref{f:ism}, together with the allowed distance ranges from
Walter (2001) and Kaplan et al.\ (2002).  The value of $N_H$ estimated
in this way at a distance of 60pc is an order of magnitude lower than
the X-ray measurement, but is in good agreement with the distance of
140pc derived by the latter authors.  This distance would place \rxj\
on the outskirts of the R~CrA cloud using the cloud distance of Knude
\& H{\o}g (1998) of $\sim 170$pc and within the cloud using
the canonical cloud distance of $\sim 130$~pc.  In either case, \rxj\ would
likely lie in a region of relatively high ISM density ($\sim 1$-10
cm$^{-3}$).  While these estimates of $N_H$ are crude and will smooth
out any small-scale ISM inhomogeneities, the larger distance is easier to
reconcile with the measured column.  The cloud distance of Knude
\& H{\o}g (1998) then represents an upper limit to the distance of
\rxj .

\section{Discussion}
\label{s:discuss}

Our results, combined with the recent analysis of Ransom et al.\
(2002), demonstrate a lack of pulsed features above a level of
2.7~\%\ (unaccelerated search; 4.5\%\ from the accelerated search of
Ransom et al.) and no unequivocal
detection of spectral features.  This dearth of indices with which to
restrict parameter space precludes an obvious answer to the problem of
the nature and origin of \rxj .

The apparent lack of electron or proton resonance cyclotron absorption
suggests that magnetic field strengths in the ranges (1-$7)\times
10^{10}$ and (0.2-$1.3)\times 10^{14}$~G are less likely, as discussed by
Paerels et al.\ (2001) for RXJ~0720.4$-$3125 and by Burwitz et al.\
(2001), but, as emphasised by the latter, should not be excluded owing
to possible difficulties in detecting the absorption features.
Indeed, neutron stars with different levels of magnetic field up to
$10^{15}$~G have now been observed with high resolution X-ray
spectrometers and none have so far shown absorption features that are
intrinsic to the stellar photosphere (e.g.\
RXJ~0720.4$-$3125---Paerels et al. 2001; PSR 0656+14---Marshall \&
Schulz 2002; Vela---Pavlov et al.\ 2001; and 4U 0142+61---Juett et al. 2002).

Our derived angular size based on modelling of the LETGS spectra,
$R_\infty/D_{100}=4.12\pm 0.68$~km/100pc, is consistent with that of
Burwitz et al.\ (2001), though the revised allowed distance range of
111-200~pc (Kaplan et al.\ 2002), together with the distance upper
limit constraint based on the R~CrA cloud, $D_{100}\leq 1.70$, now
implies a radiation radius in the range $R_\infty=3.8$-8.2~km.  The high
end of this range, corresponding to the largest allowed distance, is
still inconsistent with current ``normal'' NS equations of state 
($R_\infty\ga 12$~km; e.g.\ Lattimer \& Prakash 2000), as well as
with those with extreme softening, such as kaon condensate models.

Pons et al.\ (2001) find that heavy element-dominated atmosphere
models provide a plausible match to the low resolution ROSAT PSPC
spectra and UV and optical fluxes of \rxj , while Kaplan et al.\
(2002) alleviate conflicts with standard EOS through their revised
distance.  The heavy element models yield larger radii by virtue of
having cooler effective temperatures than blackbody spectra with
similar energy distributions.  However, Burwitz et al.\ (2001) argued
against such uniform temperature heavy element-dominated atmosphere
solutions based on a lack of the expected spectral line features.  The
apparently featureless but much higher quality LETGS spectra presented
here strengthen these conclusions.

An alternative favoured by Pons et al.\ (2001), Burwitz et al.\
(2001) and Ransom et al.\ (2002) is a non-uniform temperature
model---that we are only seeing a localised hot region on the surface
of a cooler star.  The latter authors argue that the gravitational
smearing effects described by Psaltis, {\" O}zel, \& DeDeo (2000) account for
a lack of observed pulsations.  However, the pulse fraction
expectations of Psaltis et al.\ (2000) indicate X-ray pulse fraction
levels below our 2.7~\%\ limit would be seen only $\sim 10$-15~\%\ of
the time.

Pulsation would also be expected for a young ($\sim 10^6$yr), cooling
INS with a strong magnetic field.  The alternative---that \rxj\ is an
older object re-heated by ISM accretion---has been dismissed by Kaplan
et al.\ (2002) and van Kerkwijk \& Kulkarni (2001a), largely based on
a Bondi-Hoyle accretion rate for the space velocity of Walter (2001)
and Kaplan et al.\ (2002) that would be much too low to explain the
observed luminosity, and on a possibly strong, accretion-inhibiting
magnetic field (Pons et al.\ 2001).  Nevertheless, \rxj\ appears to be
in the outskirts of the R~CrA cloud and, based on its HST-derived
velocity vector, has likely passed through more dense regions than it
now resides in.  Finding one of only a handful of INS candidates in
such a region by chance is very unlikely, yet it is just such dense
ISM regions that are expected to power accretion-heated INSs.
Based on the large velocity, however, accretion could only be
significant if the Bondi-Hoyle formalism were to be inapplicable for
\rxj .
While we tentatively ascribe the $\sim 2$\% change in observed 0th
order count rate in the 19 months separating the two observation sets
to detector effects, such variability could be accommodated by an
accretion model as the star passed through ISM density fluctuations on
$\sim$AU scales.

The lack of spectral features is also consistent with a pure hydrogen
atmosphere model that is expected to result from modest accretion,
whereby heavier elements undergo rapid gravitational settling.  Pons
et al.\ (2001) argue that standard pure H models overpredict the
optical flux of \rxj\ by a factor of 30 and that the magnetic
accreting models of Zane, Turolla \& Treves (2000), while capable of
reproducing the observed X-ray to optical flux ratio, would 
need to be two orders of magnitude brighter and an order of magnitude
hotter than observed.  However, different accretion scenarios and
atmospheric models might bear examination in light of the
otherwise coincidental location of \rxj .

The slightly unfavourable odds of a non-uniform temperature model
failing to show signs of pulsation leads us to consider a third
possibility.  Taken at face value, the distance, $N_H$, and lack of
spectral and temporal features and pulsations favour a more compact
object than current NS models permit, but one that is allowed for in
strange quark matter solutions (e.g.\ Lattimer \& Prakash 2000).
There now exists a body of evidence from heavy-ion collision
experiments supporting the viability of a quark-gluon plasma (Heinz
2001), and the possible existence of ``strange stars'' (e.g.\ Alcock
et al.\ 1986) is perhaps not as speculative as it once was.  As noted
by Pons et al.\ (2001), such an object would be expected to have a
thermal spectrum as we observe.  Such a suggestion is not
unprecedented and there now exists a small handful of objects whose
apparent compactness could be explained if they are composed of quark
matter (e.g.\ Cheng et al.\ 1998; Li et al.\ 1995; Xu et al.\ 1999).
Of the existing quark star candidates, \rxj\ arguably presents the
strongest and most direct case.  If this case survives future
scrutiny, then the likelihood of such an object being the brightest
and closest of the current few INS candidates would add some support
to speculation that such a state of matter is a common product of
supernovae explosions, or a common phase or endpoint in the evolution
of a neutron star (e.g.\ Alcock et al.\ 1986; Kapoor \& Shukre 2001;
Xu, Zhang \& Qiao 2001).

\acknowledgments

We extend warm thanks to Pete Ratzlaff for invaluable assistance and 
``wonder scripts'' developed at short notice.
We also thank members of the CfA HEAD for useful comments and
corrections that enabled us to improve the manuscript. 
The SAO authors were supported by NASA contract NAS8-39073 to the
{\em Chandra X-ray Center} during the course of this research.
HLM was supported by NASA contract SAO SV1-61010.

\clearpage


\clearpage

\clearpage

\clearpage

\begin{deluxetable}{rrrrrrr}
\tabletypesize{\small}
\tablecaption{Summary of Chandra LETG+HRC-S Observations\label{t:obs}}
\tablewidth{0pt}
\tablehead{ & & & \colhead{Exposure} & \multicolumn{2}{c}{Net
Events\tablenotemark{a}} & 
\colhead{0th Rate\tablenotemark{b}}  \\
\colhead{Obs ID} & \colhead{UT Start}   & \colhead{UT End}   &
\colhead{[s]} & \colhead{0th} & \colhead{0th + 1st} &
\colhead{[Hz]}
}
\startdata
113 & 2000-03-10 07:55:12 & 2000-03-10 23:37:24 & 55121 & 12202 & --- & 
$0.2195\pm 0.0020$\\
3382 & 2001-10-08 08:18:49 & 2001-10-09 03:01:50 & 101172 & 20949 &
86516 &
$0.2166\pm 0.0016$ \\
3380 & 2001-10-10 05:06:28 & 2001-10-12 04:00:48 & 166325 & 35097 &
135230 &
$0.2154\pm 0.0013$ \\
3381 & 2001-10-12 19:19:26 & 2001-10-14 09:14:28 & 169956 & 36011 &
141349 &
$0.2157\pm 0.0013$ \\
3399 & 2001-10-15 11:47:06 & 2001-10-15 14:42:59 & 9282 & 1962 & 7136 &
$0.2126\pm 0.0051$ \\
\enddata
\tablenotetext{a}{The net event numbers includes background events; while
including the 1st order events increases the net source events, the
number of background events also increases.}
\tablenotetext{b}{Rates were derived from 0th order data when 
valid event rates did not exceed 184~count~s$^{-1}$; this selection 
criterion yielded 417786 seconds of the 501856s total exposure time.}
\end{deluxetable}

\end{document}